\documentclass[twocolumn,10pt]{article}
\usepackage[latin1]{inputenc}
\usepackage{fullpage}
\usepackage{graphics}
\usepackage{epic}

\title{Temporal Runtime Verification using Monadic Difference Logic}
\author{Henrik Reif Andersen\\
  {\small IT University of Copenhagen}
  \and Kåre Jelling Kristoffersen\\
  {\small Lector A/S}
}
\newcommand{\atopr}[2]{{#1}({#2})}
\newcommand{\Prop}{\textit{Prop}}
\newcommand{\Pred}{\textit{Pred}}
\newcommand{\Var}{\textit{Var}}
\newcommand{\pfp}{\textit{PFP}} 
\newcommand{\nfp}{\textit{NFP}} 
\newcommand{\reals}{\ensuremath{I\!\!R_+}}
\newcommand{\nats}{\ensuremath{I\!\!N}}

\newtheorem{lem}{Lemma}

\newtheorem{cor}[lem]{Corollary}
\newtheorem{defn}{Definition}
\newenvironment{proof}{\textit{Proof:}}{\fbox{}\\}
\newcommand{\always}[1]{\texttt{always}_{#1}}
\newcommand{\eventually}[1]{\texttt{eventually}_{#1}}
\newcommand{\until}[1]{\texttt{until}_{#1}}
\newcommand{\after}[1]{\texttt{after}_{#1}}
\newcommand{\between}[2]{\texttt{between}_{#1,#2}}
\newcommand{\subst}[3]{\ensuremath{{#1}[{#3}\!\mapsto\!{#2}]}}
\newcommand{\quot}[1]{{/_{#1}}}
\newcommand{\quota}{\quot{(s,t),(s',t')}}

\begin{document}

\maketitle

\begin{abstract}
  In this paper we present an algorithm for performing runtime
  verification of a bounded temporal logic over timed runs.  The algorithm
  consists of three elements. First, the bounded temporal formula to be
  verified is translated into a monadic first-order logic over
  difference inequalities, which we call \emph{monadic difference
    logic}. Second, at each step of the timed run, the monadic difference
  formula is modified by computing a \emph{quotient} with the state and
  time of that step. Third, the resulting formula is checked for being a
  tautology or being unsatisfiable by a decision procedure for monadic
  difference logic. 

  We further provide a simple decision procedure for monadic
  difference logic based on the data structure Difference Decision
  Diagrams. The algorithm is complete in a very strong sense on a
  subclass of temporal formulae characterized as homogeneously monadic
  and it is approximate on other formulae. The approximation comes
  from the fact that not all unsatisfiable or tautological formulae
  are recognised at the earliest possible time of the runtime
  verification.


  Contrary to existing approaches, the presented algorithms do not
  work by syntactic rewriting but employ efficient decision structures
  which make them applicable in real applications within for instance
  business software.
\end{abstract}

\section{Introduction and related work}

\begin{figure}
\centerline{\resizebox{0.8\columnwidth}{!}{\includegraphics{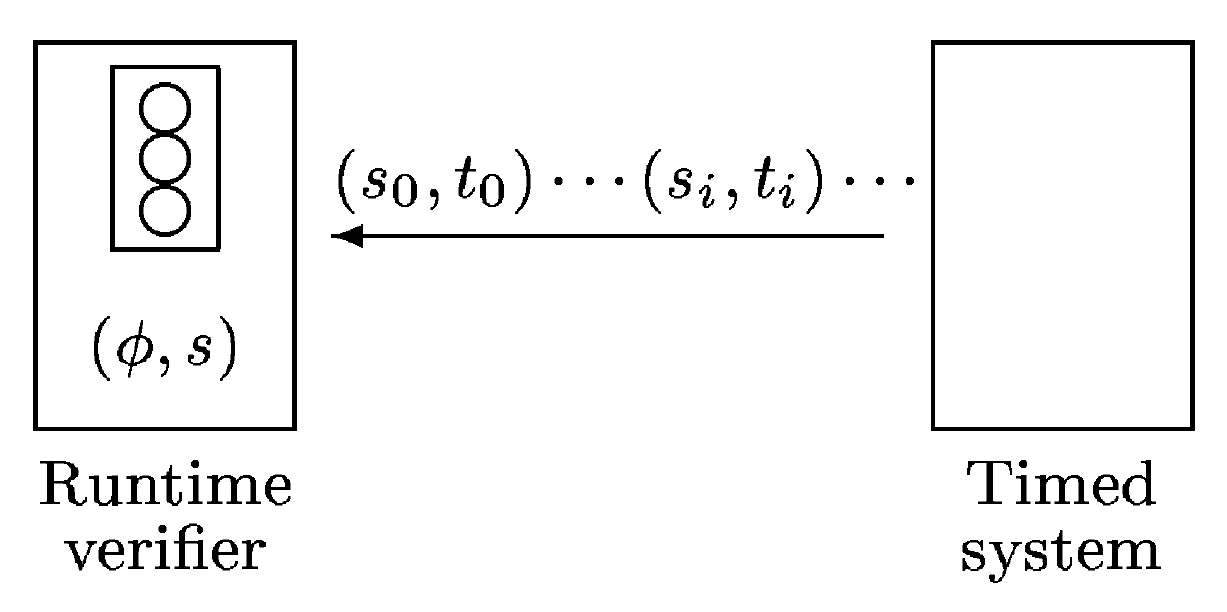}}}
\caption{The overall setup of the runtime verifier. A timed system transmits time-stamped states to the runtime verifier. The verifier is monitoring the validity of the observed run with respect to a given temporal formula $\phi$ and the last observed state $s$. The verifier has three states indicated by the traffic light: Failure, undertermined (could go either way), or acceptance (no reason to monitor any further).}\label{fig:system}
\end{figure}

Runtime verification is the task of verifying whether a running system, while it is running, satisfy given properties expressed in a suitable logic \cite{pnueli98translation,pnueli-zaks-06,havelund-et-al-ase01,kristoffersen-et-al-rv03,broy-et-al-05,drusinsky04}. Contrary to model checking the verification is not done for the complete system before running it. It is in this sense a weaker verification, since only the particular run performed by the system is checked and not all possible runs. However, it can be a much more appropriate verification since for model checking to give reliable results, a model of the environment must be supplied and if this is not correctly capturing the environment the verification becomes unreliable: it might accept or reject a property because of the existence of runs in the model which would never occur in practice. Figure \ref{fig:system} schematically illustrates the situation.

It also turns out, somehow surprisingly, that in order to perform proper runtime verification that provides \emph{timely responses} at the earliest possible time of the run, a decision problem for the logic has to be solved. For most logics, this is complexity-wise a more difficult problem than the associated model checking problem, so runtime verification might in fact be more challenging than model checking despite the apparent simplification of the problem to a particular run and not a quantification over all runs.

Existing approaches to runtime verification use invariance techniques \cite{pnueli98translation,pnueli-zaks-06,peleska96}, where a concurrent process is surveying the state of the system under verification and checks that it always obeys an invariance property, or formulae rewriting \cite{havelund-et-al-ase01,kristoffersen-et-al-rv03}, where in each step of a run the property formulae is rewritten depending on the state of the current step. 
In \cite{kristoffersen-et-al-rv03} an attempt was made to repeat the 
rewrite principle from \cite{havelund-et-al-ase01} for quantitative 
temporal logic. This attempt, however, revealed difficulty in 
expressing, within the same logic, the property which must hold for the 
remaining run after rewriting based on the current step. Moreover, 
although being complete,  no efficient methods to check a formula for 
being a tautology or unsatisfiable exists for these logics. And also, 
apparently there is no easy way to compute nearest deadlines, e.g. when 
(in the future) the current formula would become a tautology (or 
unsatisfiable) provided that the current state does not change. In this 
paper we shall take an appoach which makes these things a whole lot easier.

We take a different approach. By encoding the runtime problem as a satisfiability problem for a monadic first-order logic, we arrive at a different type of algorithm. This algorithm is capable of utilizing a powerful decision structure for difference logic which inherits some of the strengths of binary decision diagrams \cite{bryant86}. 

\section{Bounded temporal logic and monadic difference logic}

We work with two logics: a bounded temporal logic (BTL) and a monadic difference logic (MDL). We assume a set of indexed propositions $\Prop=\{p_1, p_2, \ldots\}$ and a corresponding set of indexed monadic predicates $\Pred=\{P_1, P_2, \ldots\}$. The correspondance between propositions and predicates will be exploited in the translation of the temporal logic into monadic difference logic. 

Bounded temporal logic formulae are constructed from the following grammar:
\[
\psi ::= p_j \mid \psi_1\wedge\psi_2 \mid \neg\psi \mid \texttt{always}_c \psi
\]
where $c\in\reals$ and we use $\reals$ for the set of non-negative reals. As usual there is a range of derived operators, e.g. $\texttt{eventually}_c \psi =\neg\always{c}\neg\psi$, and we use the standard definitions for $\vee$ (disjunction), $\rightarrow$ (implication), and $\leftrightarrow$ (biimplication).
The semantics of BTL is given over \emph{timed runs}. A timed run is an infinite sequence of pairs $\sigma_i=(s_i,t_i)$ of a state $s_i\subseteq{\Prop}$ and a time $t_i\in\reals$:
\[
\sigma = (s_0,t_0)(s_1,t_1)\cdots(s_i,t_i)\cdots
\]
such that $t_0=0, t_i<t_{i+1}$ for all $i\in\nats$. We call a pair
$\sigma_i=(s_i,t_i)$ a \emph{timed state}. The state $s_0$ represents
the initial state of the system. The intended interpretation of a run,
is that between two elements in the sequence, the state is
unchanged. The elements thus represent the ``events'' taking place:
an event is a state change decorated with a time stamp of when the change happens. For a
pair $\sigma_i=(s_i,t_i)$ we use the functions $s$ and $t$ for the
$s$- and $t-$components: $s(\sigma_i)=s_i,
t(\sigma_i)=t_i$. 

We further assume that all runs have \emph{finite variability} (also referred to as non-zeno runs \cite{alur94theory}) in the sense that for all $t\in\reals$, there exists $ i\in\nats$ such that $t<t_i$. This is a very reasonable assumption for timed runs coming from a running system. If the system stabilizes into no state change, redundant timed states can be generated at regular intervals.

 For $u\in\reals$ we define $\atopr{\sigma}{u}=\sigma_{i}$, where $i$ is
the largest index with $t(\sigma_i)\leq u$. For any $u\in\reals$, $\sigma(u)$ is always
well-defined because of finite variability and the fact that $t_0=0$. 
With this definition, $\atopr{\sigma}{u}$
is the timed state at time $u$ in the run $\sigma$. The state of the
system at time $u$ is $s(\sigma(u))$.

We express that a timed run $\sigma$ satisfies a BTL formulae $\psi$
at time $u$ as the relationship $\sigma\models_u\psi$ defined
inductively as follows:
\[
  \begin{array}{lll}
    \sigma\models_u p_j & \textrm{iff} & p_j\in s(\atopr{\sigma}{u})\\
    \sigma\models_u\psi_1\wedge\psi_1 & \textrm{iff} & 
       \sigma\models_u\psi_1 \textrm{ and } \sigma\models_u\psi_1\\
    \sigma\models_u\neg\psi &  \textrm{iff} & 
       \textrm{not } \sigma\models_u\psi\\
    \sigma\models_u\always{c}\psi  & \textrm{iff} & \\
    \multicolumn{3}{l}{\indent\indent
       \textrm{for all } u' \textrm{ with } u\leq u'\leq u+c, \sigma\models_{u'}\psi }\\
  \end{array}
\]
We use the abbreviation $\sigma\models\psi$ for $\sigma\models_0\psi$.

For monadic difference logic we use the presence of a set of first-order variables $\Var$ ranged over  by $x,y,z,u,v,\ldots$. Formulae in \emph{monadic difference logic} are constructed from the following grammar:
\[
\phi ::= P(x) \mid x-y\leq c \mid \phi_1\wedge\phi_2 \mid \neg\phi \mid \forall x.\phi
\]
Without the monadic predicates, this logic is known as \emph{difference logic},  separation logic \cite{cotton-et-al04,bryant-et-al-cav02,strichman-et-al-cav02}, or difference constraint expressions \cite{jm-et-al-csl99}. 
Monadic difference logic is known to have a decision procedure in PSPACE \cite{hirshfeld-et-al-fi04}.

We use the notation $\phi(x_1,\ldots,x_k;P_1,\ldots,P_l)$ for a formula with the free variables $x_1,\ldots,x_k$ and the monadic predicates $P_1,\ldots,P_l$. The semantics of a formula is then given with respect to an interpretation of the variables as reals and the monadic predicates as subsets of reals, $t_1,\ldots,t_k\in\reals, S_1,\ldots,S_l\subseteq\reals$:
\[
  (t_1,\ldots,t_k;S_1,\ldots,S_l)\models\phi(x_1,\ldots,x_k;P_1,\ldots,P_l).   
\]
The definition of satisfaction is straightforward by interpreting $x_j$ as $t_j$ and $P_j(x)$ as $x\in S_j$. We use $\models\phi$ if their exists $t_j$'s and $S_j$'s such that the above holds.

In difference logic only relative bounds of variables can be
expressed. The syntax does not allow for expressions such as $x\leq
c$. However, by introducing a special ``zero'' variable $z$, which can be
read as having constantly the value zero, we obtain a similar
expressiveness without complicating the logic.

\section{Translating BTL to MDL}
The first phase in obtaining a runtime verifier is to translate the bounded temporal logic formulae into monadic difference logic. The key ingredient is to use monadic predicates instead of propositions referring to timed states of the run. For each proposition $p_j$, we use a monadic predicate $P_j$ such that $P_j(u)$ holds if and only if $p_j\in s(\atopr{\sigma}{u})$. In terms of semantics, runs will be translated to a collection of subsets of reals. For a proposition $p_j$, and a run $\sigma$, the $j$'th set of reals $S_j$, is the set of time points for which $p_j$ holds in $\sigma$. I.e., given a run $\sigma$ the corresponding  sets are $\vec{S}(\sigma) = (S_1(\sigma),\ldots,S_k(\sigma))$, defined by:
\[
  S_j(\sigma) \ = \ \{ u\in\reals \mid  p_j\in s(\atopr{\sigma}{u}) \}.
\]
\begin{figure}
\setlength{\unitlength}{1.5pt}
\begin{picture}(150,55)(0,0)
  \put(20,23){\vector(1,0){115}}
  \put(140,23){\makebox(0,0)[l]{$t$}}
  \put(20,21){\line(0,1){4}}
  \put(60,21){\line(0,1){4}}
  \put(90,21){\line(0,1){4}}
  \put(120,21){\line(0,1){4}}
  \put(20,15){\makebox(0,0){0}}
  \put(60,15){\makebox(0,0){4}}
  \put(90,15){\makebox(0,0){7}}
  \put(120,15){\makebox(0,0){10}}
  \put(20,5){\makebox(0,0){$\{p_1\}$}}
  \put(60,5){\makebox(0,0){$\{p_1,p_2\}$}}
  \put(90,5){\makebox(0,0){$\{p_2\}$}}
  \put(120,5){\makebox(0,0){$\{p_1\}$}}
  \put(5,10){\makebox(0,0){$\sigma:$}}
  \put(20,35){\line(1,0){69}}
  \multiput(120,35)(5,0){4}{\line(1,0){2}}
  \put(60,45){\line(1,0){59}}
  \put(5,35){$S_1$}
  \put(5,45){$S_2$}
  \put(20,35){\circle*{2}}
  \put(120,35){\circle*{2}}
  \put(60,45){\circle*{2}}
  \put(90,35){\circle{2}}
  \put(120,45){\circle{2}}
\end{picture}
\caption{The relationship between a run $\sigma=(\{p_1\},0)(\{p_1,p_2\},4)(\{p_2\},7)(\{p_1\},10)$ and the corresponding monadic sets $S_1$ and $S_2$.
In this example we have, for instance, $\sigma(3)=(\{p_1\},0)$ and $\sigma(9.99)=(\{p_2\},7)$.}\label{fig:reln}
\end{figure}
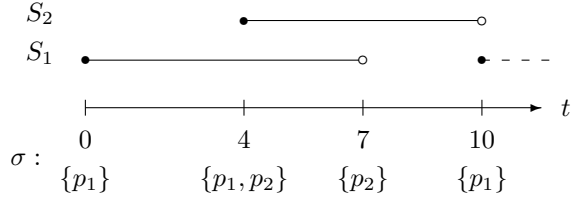
The (monadic) set $S_j$ is the semantical interpretation of the monadic predicate $P_j$.  An example of the relationship between propositions and monadic sets is shown in figure \ref{fig:reln}.

With monadic predicates, the translation is very close to being a translation into the meta-logic used in the semantics. 
The translation is defined inductively for an arbitrary ``starting point'' $x$ and goes as follows: 
\begin{eqnarray*}
T(p_j)_x &=& P_j(x)\\
T(\psi_1\wedge\psi_2)_x &=& T(\psi_1)_x \wedge T(\psi_2)_x\\
T(\neg \psi)_x &=& \neg T(\psi)_x\\
T(\texttt{always}_c\psi)_x &=& 
\forall y.\ 0\leq y-x\leq c \rightarrow T(\psi)_y
\end{eqnarray*}
For the derived operator $\texttt{eventually}_c\psi$ we obtain $T(\texttt{eventually}_c\psi)_x = \exists y. 0\leq y-x\leq c \wedge T(\psi)_y$. Observe, that $T(\psi)_x$ has only one free variable, $x$, that can be thought of as the ``starting time''.

\begin{lem}[Translation correctness]
  For all bounded temporal formulae $\psi$, timed runs $\sigma$, and time
  points $t\in\reals$, we have
  \[
  \sigma\models_t\psi,\ \textrm{ if and only if, }\ (t;\vec{S}(\sigma))\models T(\psi)_z\,.
  \]
\end{lem}
\emph{Example.} As an example consider the BTL formula:
\[
\psi = \texttt{eventually}_{8}\ \texttt{always}_3\ p_2
\]
It translates to:
\[
\begin{array}{rcl}
T(\psi)_z &=&
  \begin{array}[t]{@{}l} 
    \exists x.\ 0\leq x-z\leq 8\\
    ~~~\wedge \forall y.\ 0\leq y-x\leq 3\rightarrow P_2(y)\,.\\
  \end{array}
\end{array}
\]
Reading this, it states that there must exist a time point $x$ no more than 8 time units after $z$, such that for all time points $y$ no more than 3 time units after $x$, $P_2$ holds for $y$.

\section{Quotienting}
In runtime verification we receive one timed state $\sigma_i$ of the
run at a time. Our approach will be to translate the temporal formula
under verification, $\psi$, to a monadic formula $\phi$ using the
translation $\phi=T(\psi)_z$. After receiving each new timed state of
the run, we transform $\phi$ in order to take the additional
information into account given by the timed state. If $\sigma_i$ and
$\sigma_{i+1}$ are two consecutive timed states of the run, we form
the \emph{quotient} $\phi\quot{\sigma_i\sigma_{i+1}}$ with the property that
the state information in $\sigma_i\sigma_{i+1}$ has been taking into
account such that the resulting formula no longer refers to state in
the timing interval given by $\sigma_i$ and $\sigma_{i+1}$. Recall
that on each point in the interval $[t(\sigma_i);t(\sigma_{i+1})[$ the
state is $s(\sigma_i)$ and in the endpoint $t(\sigma_{i+1})$ the state
is $s(\sigma_{i+1})$.

For two consecutive pairs of timed states, $(s,t),(s',t')$ with $t<t'$ we
define the quotient inductively over monadic difference formulae as
shown in figure \ref{fig:quotient}.
\begin{figure*}[t]
\begin{eqnarray*}
P_j(x) \quot{(s,t)(s',t')} &= &
\left\{\begin{array}{l@{}lll}
         & (t'<x-z \wedge P_j(x)) & \textbf{if}\ p_j\not\in s, p_j\not\in s' & \mathrm{(I)}\\
x-z=t' \vee{} & (t'<x-z \wedge P_j(x)) & \textbf{if}\ p_j\not\in s, p_j\in s' & \mathrm{(II)}\\
t\leq x-z<t' \vee{} & (t'<x-z \wedge P_j(x)) & \textbf{if}\ p_j\in s, p_j\not\in s' & \mathrm{(III)}\\
t\leq x-z\leq t' \vee{} & (t'<x-z \wedge P_j(x)) & \textbf{if}\ p_j\in s, p_j\in s' & \mathrm{(IV)}
\end{array}
\right. \\
x-y \leq c  \quot{(s,t)(s',t')} &=&
x-y \leq c\\
\phi_1\wedge\phi_2 \quot{(s,t)(s',t')} &=&
(\phi_1 \quot{(s,t)(s',t')})\wedge(\phi_2 \quot{(s,t)(s',t')})\\
\neg\phi \quot{(s,t)(s',t')} &=&
\neg(\phi \quot{(s,t)(s',t')})\\
\exists x.\phi  \quot{(s,t)(s',t')} &=&
\exists x.(\phi  \quot{(s,t)(s',t')})
\end{eqnarray*}
\caption{Quotienting of monadic difference formulae over a pair of timed states $(s,t),(s',t')$ with $t<t'$.}\label{fig:quotient}
\end{figure*}

The quotient distributes over all operators and make only a change to
the formula at the point when a monadic predicate is met. In fact, the
quotient could also be viewed as simply the substitution
\[\subst{\phi}{P_j(x) \quota}{P_j(x)}_{j=1}^l\]
 on all predicates $P_j(x)$.  In order to formally state the relevant
properties of the quotient, we use the notion of two subsets of
$\reals$ agreeing on another subset: The sets $S,S'\subseteq\reals$
\emph{agree on} $D\subseteq\reals$ if $S\cap D=S'\cap D$, i.e.,
$\forall t\in D. t\in S\Leftrightarrow t\in S'$. Two collections of sets $\vec{S},\vec{S'}$ pairwise agree on $D$ if for all $j=1,\ldots,l$, $S_j$ and $S_j'$ agree on $D$. A monadic difference formula $\phi$ is \emph{independent of states} on $C\subseteq\reals$ if, for all $\vec{S},\vec{S'}$ that pairwise agree on $D=\reals\setminus C$, we have that for all $\vec{t}\in\reals^k$:
  \[
    (\vec{t};\vec{S})\models \phi,\ \textrm{ if and only if, }\ 
    (\vec{t};\vec{S'})\models \phi\,.
  \]
Further, we say that
$(s,t)(s',t')$ \emph{is consistent with $S_j\subseteq\reals$}, if
\[
\begin{array}{l}
  (t'\in S_j\Leftrightarrow p_j\in s'),\ \textrm{and}\\ 
  \textrm{for all}\ t''.\ 
    t\leq t'' <t' \Rightarrow
    (t''\in S_j\Leftrightarrow p_j\in s)
\end{array}
\]
We can now formally state the properties of the quotient in the following lemma:
\begin{lem}[Quotienting lemma]
  Let $(s,t),(s',t')$ be pairs of timed states with $t<t'$ and $\phi$
  a monadic difference formula with one free variable.\\
  \textbf{(Independence)} The quotient $\phi\quota$ is independent of the states on the interval $[t;t']$.\\
  \textbf{(Correctness)} If $(s,t)(s',t')$ is consistent with $S_j$ for all $1\leq j\leq l$ then for all ${t''}\in\reals^k$ we have:
  \[\begin{array}{l}
    ({t''};\vec{S})\models \phi\ \textrm{ iff }\ 
    ({t''};\vec{S})\models \phi\quota\,.
   \end{array}
  \]
  \textbf{(Preservation)} If $t<t'$ and $\phi$ is independent of the states on $[t;t']$ then $\phi\quot{(s',t')(s'',t'')}$ is independent of the states on $[t;t'']$. 
\end{lem}
Given a timed run $\sigma$ with the finite prefix $\sigma^{0-i}=\sigma_0\sigma_1\cdots\sigma_i$ of its first $i+1\geq 2$ timed states, we denote by $\phi\quot{\sigma^{0-i}}$ the repeated quotient $\phi\quot{\sigma_0\sigma_1}\quot{\sigma_1\sigma_2}\quot{\cdots}\quot{\sigma_{i-1}\sigma_i}$. 
From the independence and preservation properties, it follows that $\phi\quot{\sigma^{0-i}}$ is independent of states on $[0;t(\sigma_i)]$. In order words the formula $\phi$ has been modified to reflect the past and its validity now only depends on the future. If $\phi\quot{\sigma^{0-i}}$ is a tautology, then no matter what timed states will occur in the future, we know that $\phi$ is going to hold for all runs. Similarly, if $\phi\quot{\sigma^{0-i}}$ is unsatisfiable, no possible future run will be able to make $\phi$ become fulfilled. Checking for these two situations is the main part of our first runtime verification algorithm.

\medskip

\emph{Example (continued).}
Consider again the temporal formula $\psi = \texttt{eventually}_{8}\ \texttt{always}_3\ p_2$ with translation 
$T(\psi)_z=\exists x.\ 0\leq x-z\leq 8 \wedge \forall y.\ 0\leq y-x\leq 3\rightarrow P_2(y)$. The quotient $T(\psi)_z\quot{(\{p_1\},0)(\{p_1,p_2\},4)}$ can be computed to be
\[
\begin{array}{l}
\exists x.\ 0\leq x-z\leq 8 \wedge \forall y.\ 0\leq y-x\leq 3\rightarrow \\
~~~~~~(P_2(y)\quot{(\{p_1\},0)(\{p_1,p_2\},4)}),
\end{array}
\]
using the cases for $\exists$, $\wedge$ etc. of the quotienting of figure \ref{fig:quotient}. Furthermore, we have according to case (II) of the quotienting on monadic predicates in figure \ref{fig:quotient}, that $P_2(y)\quot{(\{p_1\},0)(\{p_1,p_2\},4)}$ becomes
\[
{y-z=4} \vee (4 < y - z \wedge P_2(y))\,.
\]
This gives the combined result:
\[
\begin{array}{l}
\exists x.\ 0\leq x-z\leq 8 \wedge \forall y.\ 0\leq y-x\leq 3\rightarrow \\
~~~~~~{y-z=4} \vee (4 < y - z \wedge P_2(y))\,.
\end{array}
\]
We call this expression $\phi$ and can now compute the next quotient $\phi\quot{(\{p_1,p_2\},4)(\{p_2\},7)}$. Using the various cases of the quotienting of figure \ref{fig:quotient}, in particular case (IV), it is not hard to see that we end with:
\[
\begin{array}{l}
\exists x.\ 0\leq x-z\leq 8 \wedge \forall y.\ 0\leq y-x\leq 3\rightarrow \\
~~~~~~{y-z=4} \vee 4 < y - z \leq 7 \vee {}\\
~~~~~~(7 < y-z\wedge P_2(y))\,.
\end{array}
\]
This can be simplified to:
\[
\begin{array}{l}
\exists x.\ 0\leq x-z\leq 8 \wedge \forall y.\ 0\leq y-x\leq 3\rightarrow \\
~~~~~~4 \leq y - z \leq 7 \vee (7 < y-z\wedge P_2(y))\,.
\end{array}
\]

Notice, that taking $x=4$ the universal quantification becomes true irrespectively of the monadic predicate $P_2(y)$. Therefore, this expression is a tautology and it would be safe for a runtime verifier to conclude at time 7 that the formula is fulfilled.

\section{Runtime verification algorithms: MDLV and DLV}
In \cite{hirshfeld-et-al-fi04} a PSPACE decision algorithm for monadic difference logic is described. In the runtime verification algorithm we use isTautMDL($\phi$) to denote a run of this algorithm to check for tautologiness, and isUnsatMDL($\phi$) to denote a run checking for unsatisfiability. The general checking algorithm is shown in figure \ref{fig:mdl-alg}.

\begin{figure}
\begin{quote}
\begin{tabbing}
XXXX\=XX\=\kill
1: MDLV($\psi$)\\
2: \> $\phi$ := $T(\psi)_0$\\
3:\> $t$ := $0$\\
4: \> $s$ := $s_0$ /* initial state at time 0 */\\
5: \> \textbf{while not} isTautMDL($\phi$) \\
5a:\> \> \textbf{and not} isUnsatMDL($\phi$)\\
6: \> \> wait for next timed state $(s',t')$\\
7: \> \> $\phi$ := $\phi \quota$\\
8: \> \> $s$ := $s'$\\
9: \> \> $t$ := $t'$\\
10: \> \textbf{end}\\
11: \> \textbf{if} isTautMDL($\phi$) \textbf{then}\\
11a:\>\> $\psi$ is already fulfilled by current run\\
12: \> \textbf{else} $\psi$ will never be fulfilled by\\
12a:\>\> continuing the current run
\end{tabbing}
\end{quote}
\caption{MDLV: Monadic Difference Logic Verifier. A runtime verification algorithm using a general decision procedure for monadic difference logic. It is sound and complete.}
\label{fig:mdl-alg}
\end{figure}

Although correct and complete, the algorithm is going to be rather impractical because of the complicated decision procedure for MDL \cite{hirshfeld-et-al-fi04}. Instead, we develop a much more efficient algorithm utilizing Difference Decision Diagrams \cite{jm-et-al-csl99,jm-dddlib,MLAH1999smc,jm-jlap02,jm-apn02,jm-phd}. This algorithm is going to use properties of formulae being monotonic in the monadic predicates. To make this more explicit, we assume that the monadic difference formula is converted to \emph{positive form} by requiring it to be expressed in the following restricted grammar:
\[
\phi ::= 
\begin{array}[t]{l}
P(x) \mid \neg  P(x) \mid x-y\leq c \mid x-y\not\leq c \mid\\
\phi_1\wedge\phi_2 \mid  \phi_1\vee\phi_2 \mid \forall x.\phi \mid \exists x.\phi\,.
\end{array}
\]
A formula is converted to positive form by pushing negations down through the operands, dualizing the operands, and continue until a negation gets absorped by another negation or hits a monadic predicate or inequality. We collectively refer to $P(x)$ and $\neg P(x)$ as the \emph{literal predicates} and name them as $L_i, i=1,\ldots,2k$ taking  $L_i=P_i, i=1, \ldots, k$, $L_{k+i}=\neg P_i, i=1,\ldots,k$. 

For a formula $\phi$ in positive form we denote by $\hat\phi$ the \emph{literal version}, where the $L_i$'s are used explicitly in place of the literal predicates. For instance, if $\phi=P_1(x)\wedge(\neg P_2(y)\vee\neg P_1(y))$ then $\hat\phi=L_1(x)\wedge(L_{k+2}(y)\vee L_{k+1}(y))$. Notice, that $\hat\phi$ enjoys a particular monotonicity property of its literals: the more reals on which they hold, the ``more valid'' the formula becomes. To make this precise, we first extend the subset-ordering on sets of reals pointwise to $k$-collections of reals: $\vec{S}\subseteq\vec{S'}$ if and only if for all $j=1,\ldots,2k$ we have $S_j\subseteq S'_j$.  

\begin{lem}
  If $\vec{S},\vec{S'}$ are two $2l$-collections of subsets of
  $\reals$ with $\vec{S}\subseteq\vec{S'}$, then for any $\phi$ in
  positive form with predicates among $P_1,\ldots, P_l$ and
  corresponding literal version $\hat\phi$ we have:
  \begin{itemize}
    \item If $(\vec{t};\vec{S})\models \hat\phi$ then 
      $(\vec{t};\vec{S'})\models \hat\phi$.
    \item If  $(\vec{t};\vec{S'})\not\models \hat\phi$ then
      $(\vec{t};\vec{S})\not\models \hat\phi$.
  \end{itemize}
\end{lem}
As an immediate corollary, we get a method for reducing
tautology-checking for MDL to difference logic by replacing literal predicates with the constants $0$ (for falsehood) and $1$ (for truth):
\begin{cor}\label{cor:mono1}
  Let $\phi$ be a monadic difference formula in positive form, with
  predicates among $P_1,\ldots, P_l$. Take
  $\phi^0=\hat\phi[0/L_i]_{i=1}^{2l}$ and
  $\phi^1=\hat\phi[1/L_i]_{i=1}^{2l}$ then we have:
  \begin{itemize}
    \item If $\phi^0$ is a tautology, then  $\phi$ is a tautology.
    \item If $\phi^1$ is unsatisfiable, then $\phi$ is unsatisfiable.
  \end{itemize}
\end{cor}
We can now use simpler decision procedures because we can safely replace the monadic predicates with constants. This gives rise to the algorithm in figure \ref{fig:dlv} where we used DDD-based decision procedures. DDDs are introduced in the next section.

Of course, the method in general only provides a sufficient test for
tautologiness, and is not complete as we might have that
$\phi$ is a tautology without $\phi^0$ being one. A simple example is
$\phi=\forall x.\ P_1(x) \vee \neg P_1(x)$, which is clearly a
tautology, but $\phi^0=\forall x.\ 0\vee 0=0$ is not. (In fact, $\phi^0$ is
unsatisfiable in this case.)
One way of thinking about the substitution of false respectively true for the literal predicates is as that of ``assuming the least about the future''. Although, this might seem a rather crude approximation, it will turn out that for an interesting class of properties, the algorithm is in fact going to be complete. First, however, we introduce the efficient underlying data structure of DDDs.

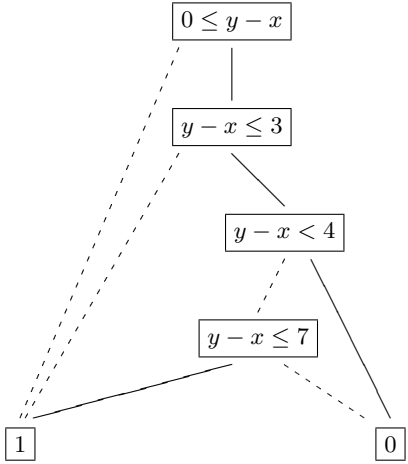
\begin{figure}[thbp]
\small
\begin{picture}(180,180)(-10,100)
  \put(100,270){\makebox(0,0){\framebox{$0\leq y-x$}}}
  \put(100,260){\line(0,-1){20}}
  \dashline{2}(80,260)(20,120)
  \put(100,230){\makebox(0,0){\framebox{$y-x\leq 3$}}}
  \put(100,220){\line(1,-1){20}}
  \dashline{2}(80,220)(22,120)
  \put(120,190){\makebox(0,0){\framebox{$y-x<4$}}}
  \put(130,180){\line(1,-2){30}}
  \dashline{2}(120,180)(110,160)
  \put(110,150){\makebox(0,0){\framebox{$y-x\leq 7$}}}
  \drawline(100,140)(25,120)
  \dashline{2}(120,140)(150,120)
  \put(160,110){\makebox(0,0){\framebox{$0$}}}
  \put(20,110){\makebox(0,0){\framebox{$1$}}}
\end{picture}
\caption{An example of a DDD representing the solutions to the difference inequality expression $0\leq y-x\leq 3\rightarrow 4 \leq y - z \leq 7$.  All edges are directed downwards. In this DDD all paths are feasible (i.e., every path is traversed by some assignment.)}\label{fig:ddd}
\end{figure}

\section{DDDs}
Difference Decision Diagrams (DDDs) \cite{jm-et-al-csl99} is a data structure for representing sets of spaces as defined by difference inequalities. More precisely, they are used for manipulating spaces defined by the following little grammar for difference logic (without monadic predicates):
\[
\phi ::= 
\begin{array}[t]{l}
x-y\leq c  \mid \neg\phi\mid\phi_1\wedge\phi_2 \mid \forall x.\phi\,.
\end{array}
\]
DDDs is an extension of Binary Decision Diagrams \cite{bryant86}, using an annotation of nodes with difference inequalities instead of Boolean variables. Figure \ref{fig:ddd} shows the DDD for $0\leq y-x\leq 3\rightarrow 4 \leq y - z \leq 7$ (the body of the resulting expression in the example above with $0$ substituted for $P_2(y)$). The DDD is read as follows: For a given set of assigment of values to the variables $x,y,z$, the expressions are evaluated in the nodes starting from the root. If a node evaluates to true the solid edge is followed to the next node. If a node evaluates to false the dashed edge is followed. If eventually the terminal node $1$ is reached, the assignment belongs to the set represented by the DDD, and if $0$ is reached it does not. 

DDDs are not canonical and atypical DDD will contain infeasible paths, i.e., paths traversed by no assignments. However, there are efficient heuristic algorithms for testing tautologiness and satisfiability as well as for realizing the logical operators as manipulations of the data structure. Details can be found in \cite{jm-et-al-csl99} and \cite{jm-phd}.


\begin{figure}
\begin{quote}
\begin{tabbing}
XXXX\=XX\=\kill
1: DLV($\psi$)\\
2:\> $\phi$ := positive form of $T(\psi)_z$\\
3: \> $t$ := $0$\\
4: \> $s$ := $s_0$ /* initial state at time 0 */\\
5: \> $\phi^0$ := $\phi[0/L_i]_{i=1}^{2k}$\\
6: \> $\phi^1$ := $\phi[1/L_i]_{i=1}^{2k}$\\
7: \> \textbf{while not} isTautDDD($\phi^0$)\\
7a: \>\> \textbf{and not} isUnsatDDD($\phi^1$)\\
8: \> \> wait for next $(s',t')$\\
9: \> \> $\phi$ := $\phi \quota$\\
9a: \>\> ~~~~ /* $\phi$ remains in positive form */\\
10: \> \> $t$ := $t'$\\
11: \> \> $s$ := $s'$\\
12: \>\> $\phi^0$ := $\phi[0/L_i]_{i=1}^{2k}$\\
13: \> \> $\phi^1$ := $\phi[1/L_i]_{i=1}^{2k}$\\
14: \> \textbf{end}\\
15: \> \textbf{if} isTautDL($\phi^0$) \textbf{then}\\
15a: \>\> $\psi$ is already fulfilled by current run\\
16: \> \textbf{else} $\psi$ will never be fulfilled by\\
16a: \>\> continuing the current run
\end{tabbing}
\end{quote}
\caption{DLV: Difference Logic Verifier. A runtime verification
algorithm using the two decision procedures for difference logic based
on DDDs: isTautDDD and isUnsatDDD.}
\label{fig:dlv}
\end{figure}

Using isTautDDD and isUnsatDDD we obtain soundness, i.e., if DLV determines fulfillment or non-fulfillment of the formula on a run, this is correct. However, there is no general guarantee on always reaching this decision.
\begin{lem}[DLV Soundness]
  DLV is sound.
\end{lem}
An example of a somewhat stupid formula for which MDLV will give
the correct answer immediately, but DLV not until time point $c$
is the following:
\[
  \eventually{c} p \wedge\always{c}\neg p\,.
\]
Recalling the definition of the eventuality modality, this formula is equivalent to $(\neg \always{c}\neg p)\wedge\always{c}\neg p$, which through propositional reasoning is clearly unsatisfiable. When translated into monadic difference logic there will be a $P$ and a $\neg P$, which are both replaced by the constant 0 or 1 in the DLV algorithm. Although this example is so simple it is easy to see how to fix it, this is not an easy task in general. But for a special class of formulae DLV is complete in a very strong sense.

\section{Homogeneously monadic formulae and completeness}

We will consider a large interesting subclass of formulae for which DLV is complete in a precise timely manner to be defined. First however we need to introduce the concept of homogenenously monadic formulae. For this we use $\pfp(\phi)$ to denote the set of predicates appearing positively in $\phi$, not under any negation, and $\nfp(\phi)$ to denote the set of predicates appearing inside a negation in $\phi$.
\begin{defn}
  A monadic difference formula in positive form $\phi$ is
  \emph{homogeneously monadic} if all predicates appear consistently
  in positive or negative form in $\phi$, i.e., 
  $\pfp(\phi)\cap\nfp(\phi)=\emptyset$.
\end{defn}

\begin{lem}
\textrm{\textbf{(Completeness for homogeneously monadic formulae)}\label{lem:homo-completeness}}
  Assume $\phi$ is a homogeneously monadic formula. Let
  $\phi^{0}=\hat\phi[0/L_i]_{i=1}^{2l}$ and
  $\phi^{1}=\hat\phi[1/L_i]_{i=1}^{2l}$. We then have:
  \begin{itemize}
    \item $\phi^0$ is a tautology, if and only if, $\phi$ is a tautology,
    \item $\phi^1$ is unsatisfiable, if and only if, $\phi$ is unsatisfiable,
  \end{itemize}
\end{lem}
\begin{proof}
  The only if directions follow from corollary \ref{cor:mono1}. For
  the other direction, assume first that $\phi$ is a tautology. This
  means that for all $l$-collections of subsets of reals $\vec{S}$,
  and $k$-vector of reals $\vec{t}\in\reals^k$, we have
  $(\vec{t};\vec{S})\models\phi$. In particular, it is valid for the
  collection $\vec{S}$ with $S_j=\emptyset$ when $P_j$ occurs (only)
  positively in $\phi$, and $S_j=\reals$ when $P_j$ occurs only
  negatively or not at all in $\phi$. Construct now a $2l$-collections
  of sets $\vec{S}'$ with each entry equal to $\emptyset$. It is now
  clear that $\hat\phi$ evaluates on $\vec{S}'$ to the same value as
  $\phi$ on $\vec{S}$. Therefore, also $\phi^0$ is a tautology.

  The case for unsatisfiability follows the same (dual) arguments.
\end{proof}

\noindent
A range of common type of formulae are indeed homogeneously
monadic. Examples are:
\begin{enumerate}
  \item \textbf{``Leads to''}:\\ $T(\always{} (p_1 \rightarrow
  \eventually{c}\neg p_1))_z$. The proposition $p_1$ appears both
  positively and negatively but the translated monadic formulae in
  positive form $\forall x.\ 0 > x-z \vee (\neg P_1(x)\vee\exists
  y. 0\leq y-x\leq 30\wedge\neg P_1(x))$, has $P_1$ only appearing
  negatively.
  \item \textbf{``Always eventually''}:\\ $T(\always{c}\eventually{d}
  p_1)_z$. Contains only one occurence of $P_1$ and is therefore
  trivially homogeneously monadic.
  \item \textbf{``Eventually always''}:\\ $T(\eventually{c}\always{d}
  p_1)_z$. Contains also only one occurence of $P_1$.  
\end{enumerate}
An example of a non-homogeneously monadic formula is   $T(\always{c}p_1\wedge\eventually{c}\neg p_1)_z$.

For a practical application such as alarms and alerts in business software it is highly desirable that a violation or fulfillment of a temporal formula is detected \emph{in time}. In time can be interpreted as the earliest possible time $t$ for which the formula, given the current run, is doomed to result in acceptance or rejection. 

Two different properties on a runtime verification algorithm could be
applied here. First, if the algorithm upon receiving a timed state,
which enforces the formula to be a tautology or unsatisfiable, detects
this immediately, we consider it to be \emph{(externally timely)
complete}. Second, if the algorithm further is capable of computing
the next earliest time-point where, if the system does not change
state before this time, the formula is doomed to become a tautology or unsatisfiable, we consider it to be \emph{internally timely complete}.  This last property could be used to warn about future failures (or successes): If the state does not change before the next unsatisfiability timepoint, the rule fails.

From lemma \ref{lem:homo-completeness}, the following corrollary immediately follows:
\begin{cor}
  DLV is externally timely complete for homogeneously monadic formulae.
\end{cor}


\begin{figure}
\centerline{\resizebox{0.8\columnwidth}{!}{\includegraphics{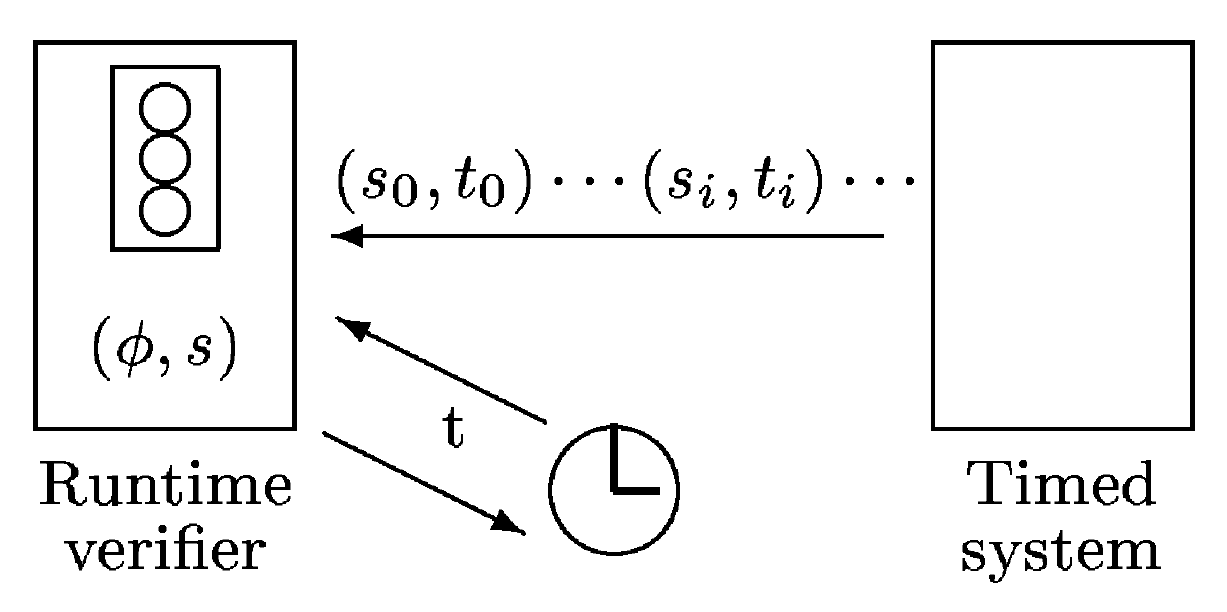}}}
\caption{The overall setup of the runtime verifier with a timer in order to obtain timely announcements of tautologiness and unsatisfiability.}\label{fig:system-with-timer}
\end{figure}


The property of being internally timely complete is harder to
obtain. Let us first be precise about the required timepoint. Let the
\emph{earliest tautology timepoint from $t$} abbreviated
ETT($\phi,s,t$) be the time $t'$, which is the earliest time $t'\geq t$
for which the formula $\phi$ becomes a tautology without changing
the state, i.e., the smallest $t'\geq t$ for which $\phi\quota$ is a
tautology. If $\phi$ is already a tautology, $t'=t$. If no such $t'$ exists, we take ETT to be $\infty$.  Let EUT($\phi,t$) similarly be the
\emph{earliest unsatisfiability timepoint from $t$}, i.e. the
smallest $t'\geq t$ for which $\phi\quota$ is unsatisfiable if
such a $t'$ exists and $\infty$ otherwise.

\emph{Example.} The formula $T(\eventually{10}p_1)_0$ on state $(\emptyset,0)$ has ETT $\infty$: A state change is required to make it a tautology. It has EUT $10$. The formula $T(\always{10}p_1)_0$ on state $(\{p_j\},0)$ has ETT $10$. It has EUT $\infty$.

Being able to compute ETT and EUT is stronger than being able to compute tautologiness and unsatisfiability because tautologiness follows from ETT if ETT is the current timepoint and similary for EUT and unsatisfibility.

In order to make an algorithm such as DLV internally timely
complete, we compute $t'=\textrm{ET}(\phi,s,t)=\min\{
\textrm{ETT}(\phi,s,t), \textrm{EUT}(\phi,s,t) \}$ and ``inject'' an
extra timed state at this timepoint, if $t'$ is not $\infty$ and no other
timed state is received before $t'$. It amounts to replacing line 8 of the algorithm in figure \ref{fig:dlv} with the following:

\begin{quote}
\begin{tabbing}
XX\=\=\kill
8a: \> \> compute $t_{et}=\textrm{ET}(\phi,s,t)$\\
8b: \> \> wait for the first of\\
\>\>~~~~ next $(s',t')$ and time-point $t_{et}$\\
8c: \> \> if time-point $t_{et}$ is reached before\\
\>\>~~~~ next state received take $t'=t_{et},s'=s$\\
\end{tabbing}
\end{quote}
The weakest unsatisfiability and tautology time-points might provide
very interesting information in themselves. For instance, the earliest unsatisfiability timepoint indicates, when, if nothing happens, at what time the next property will
fail. We leave it as an open question to find general algorithms for computing
ETT and EUP. 

\section{Other modalities}

The approach shown in this paper work for all temporal operators for which a translation to MDL is possible. There is for instance no problem adding these operators:
\[
  \psi ::=\begin{array}[t]{l}
  \cdots \mid \always{}\,\psi   \mid
  \after{c}\psi\mid\\ \between{c}{d}\psi \mid \psi_1\until{=c}\psi_2\mid\\ \psi_1\until{}\psi_2
  \end{array}
\]
with the translations:
\begin{eqnarray*}
  T(\always{}\, \psi)_x &=& \forall y. 0\leq y-x \rightarrow T(\psi)_y\\
  T(\after{c}\, \psi)_x &=& \exists y. c\leq y-x \rightarrow T(\psi)_y\\
  T(\between{c}{d}\, \psi)_x &=& \exists y. c\leq y-x\leq d \rightarrow T(\psi)_y
\end{eqnarray*}

\[\begin{array}{l}
  T(\psi_1\until{=c}\psi_2)_x =\\
\indent
  (\forall y. 0\leq y-x < c \rightarrow T(\psi_1)_y) \wedge{}\\ 
\indent
  (\forall u. u-x=c\rightarrow T(\psi_2)_u)
\end{array}
\]
\[
\begin{array}{l}
  T(\psi_1\until{}\psi_2)_x =\\
\indent
 \exists y. (0\leq y-x \wedge T(\psi_2)_y \wedge{}\\
\indent~~~~ \forall u. (0\leq u-x\wedge u-y<0)\rightarrow T(\psi_1)_u)
\end{array}
\]


\section{Conclusion and future work}
We have shown how to implement real-time runtime verification with an algorithm based on Differece Decision Diagrams as the basis of decision procedures for difference logic. Of course, other for instance SAT-based solvers could be replaced for DDDs. The key step we show, is the reduction from the runtime verification problem to a simpler decision problem on difference logic.

We are currently implementing the DDD-based algorithms and will publish reports on the results elsewhere. A first running implementation was carried out in \cite{sudhakar}. Performance should be established on real data from for instance a business software applicatoin. A discussion on an architecture that would allow a runtime verifier as the one presented in this paper to be applied to business software is discussed in \cite{ky-arch-05}.

An interesting path to take is to work directly with the monadic difference logic in formulating properties of real systems. The algorithm work for the full logic, the question is to what extent it is easy and natural to formulate real properties in the logic.

\bibliographystyle{plain}
\bibliography{rtv-mdl}

\end{document}